# CBM-Dual: A 65-nm Fully Connected Chaotic Boltzmann Machine Processor for Dual Function Simulated Annealing and Reservoir Computing

Kanta Yoshioka[1], Soshi Hirayae[1], Yuichiro Tanaka[1], Yuichi Katori[2], Takashi Morie[1], Hakaru Tamukoh[1]
[1]Kyushu Institute of Technology, Kitakyushu, Japan; [2]Future University Hakodate, Hakodate, Japan

## Abstract

This paper presents CBM-Dual, the first silicon-proven digital chaotic dynamics processor (CDP) supporting both simulated annealing (SA) and reservoir computing (RC). CBM-Dual enables real-time decision-making and lightweight adaptation for autonomous Edge AI, employing the largest-scale fully connected 1024-neuron chaotic Boltzmann machine (CBM). To address the high computational and area costs of digital CDPs, we propose: 1) a CBM-specific scheduler that exploits an inherently low neuron flip rate to reduce multiply-accumulate operations by 99%, and 2) an efficient multiply splitting scheme that reduces the area by 59%. Fabricated in 65nm ($12mm^2$), CBM-Dual achieves simultaneous heterogeneous task execution and state-of-the-art energy efficiency, delivering ×25–54 and ×4.5 improvements in the SA and RC fields, respectively.

## Introduction

Realizing autonomous-edge AI, particularly Physical AI systems, requires capabilities that complement Generative AI, including real-time decision-making under strict constraints and intelligent recognition through lightweight environment adaptation. Chaotic dynamics (CD) provides a unified hardware solution for these requirements. CD enables efficient solution searches using a chaotic simulated annealing (SA) strategy for combinatorial optimization [1] and provides powerful nonlinear transformations for time-series data processing in reservoir computing (RC) [2] (Fig. 1). To fully exploit the potential of CD, a fully connected topology of CD processors (CDPs) is essential. In the SA field, it enables the direct mapping of quadratic combinatorial optimization problems without graph embedding overhead, whereas in the RC field, it facilitates the generation of rich and diverse CD. However, fully connected CDPs face two limitations: 1) the massive number of multiply-accumulate (MAC) operations, which scales quadratically with the number of neurons [3], and 2) the high silicon area cost of high-precision multipliers for precise CD control [3].

This paper presents CBM-Dual, a digital fully connected CDP that addresses these issues using two proposed methods. CBM-Dual supports both SA and RC based on a chaotic Boltzmann machine (CBM) [4]. As shown in Fig. 2, full connectivity allows neurons to be dynamically partitioned into arbitrary number of SA and RC clusters via flexible connectivity, enabling simultaneous execution of heterogeneous tasks without area or power overheads, such as those required for autonomous robots.

## Proposed CBM-Dual Architecture

Fig. 3 shows the architecture comprising fully connected 1024-neuron CBM units, input/output layer units, and a scheduler. After initialization (Fig. 4), in the RC, 8-bit inputs are converted to 256-step pulses. These pulses are used to calculate $Z_{i,t}$ in the only RC-assigned CBM neurons, whose states are fed into the output layer unit to calculate RC outputs. The SA-assigned neurons states are output directly as SA solutions. The CBM and output layer MAC operations are controlled by a scheduler with maximized pipeline utilization.

To address the massive computational costs, the first proposed method is a CBM-specific scheduler (Fig. 5). The scheduler controls the MAC operations for $Z_{i,t}$ and $O_{i,t}$ by monitoring the neuron flips, using a delta-driven multiply accumulation (DDMAC) scheduling scheme [5]. This scheme skips unnecessary calculations by processing only flipped neurons. The effectiveness of the DDMAC scheme depends on neuron dynamics. For example, in conventional stochastic SA strategies [5, 6], the average neuron flip rate is several tens of percent, limiting the speed-up rate, even when SA chips [5, 6] employ the DDMAC scheme. However, by following a chaotic deterministic trajectory, the CBM suppresses unnecessary transitions in the search space [1]. Consequently, the CBM achieves a remarkably low flip rate of only 1% and a 99% MAC operation reduction. Furthermore, the low CBM flip rate was also confirmed in practical traffic flow optimization. The overlooked inherently low flip rate of the CBM enhances the effectiveness of the DDMAC scheme and significantly reduces the overall computation time of CBM-Dual.

To address the large area cost, the second proposed method is an adaptive temperature multiply splitting (ATMS) scheme (Fig. 6). At step $t$, the $i$-th CBM neuron's internal state $X_{i,t}$ and external state $S_{i,t}$ are updated using $Z_{i,t}$. To reduce the area overhead of 19-bit multipliers for the scaling, with temperature parameter $T$, $Z_{i,t}/T$, ATMS reformulates the operations as $(Z_{i,t} \times \alpha)/T_0$ by defining $T = T_0/\alpha$ ($T_0$: power of 2, 6-bit $\alpha$ >1). This replaces the large multiplier with a barrel shifter and a compact 6-bit multiplier. Since $S_{i,t}$ flips deterministically when $(1 - 2S_{i,t-1}) \times Z_{i,t}/T_0 > 8$ regardless of $\alpha$, precise calculation is required only in a limited range. This scheme reduces the silicon area to ×0.41 without a loss of accuracy.

## Measurement Results and Conclusion

Measurements of the 65nm $12mm^2$ CBM-Dual chip (Figs. 7, 8) show 350MHz operation at 1.2V, with peak power values of 876mW and 888mW. In Fig.7, CBM-Dual solved sparse and fully connected ($K_{1000}$) 1000 neuron max-cut problems in 3.02ms and 4.42ms, respectively, achieving an average flip rate of approximately 1%. Although the chaotic SA strategy requires more search steps than stochastic SA strategies [5, 6], the CBM-specific scheduler reduces MAC operations by 99%. The total number of MAC cycles required to reach GW-SDP score was approximately 1/2 of [6] and 1/8 of [5].

In Fig.8, CBM-Dual RC performance (memory capacity and nonlinearity) was evaluated using short-term memory (STM) and parity-check (PC) tasks. CBM-Dual achieved higher scores on both STM and PC tasks than the previous RC [7-10]. Furthermore, CBM-Dual achieved state-of-the-art accuracy on NARMA10 task with a processing time of 11μs per data point, outperforming RC on both GPU and CPU.

Simultaneous execution with 1024 neurons was partitioned into a 500-neuron SA and 524-neuron RC, executing a max-cut problem and the NARMA10 task. These results confirm the ability of CBM-Dual to manage multiple tasks simultaneously.

Tables I and Table II compare CBM-Dual with state-of-the-art processors. In the SA field, CBM-Dual is the first silicon-digital annealer that employs a chaotic strategy. In contrast to the task-specific 1024-neuron processor in [11], CBM-Dual supports arbitrary weight mapping, making it the largest-scale fully connected SA processor. The Energy efficiency per step was enhanced by ×25–54 and energy-to-solution by ×2–14. Moreover, CBM-Dual is the first silicon-digital RC processor that achieves the highest NARMA10 accuracy with the lowest latency and ×4.5 better energy efficiency. Fig. 9 shows the measurement setup and summary. We demonstrated real-time hand gesture detection through lightweight adaptation to individual characteristics, as a function for Physical AI.

**Acknowledgement** This research is based on results obtained from a project, JPNP16007, commissioned by the New Energy and Industrial Technology Development Organization (NEDO). This work receives support from JSPS KAKENHI 23H03468 and 24KJ1820, JST ALCA-Next JPMJAN23F3.

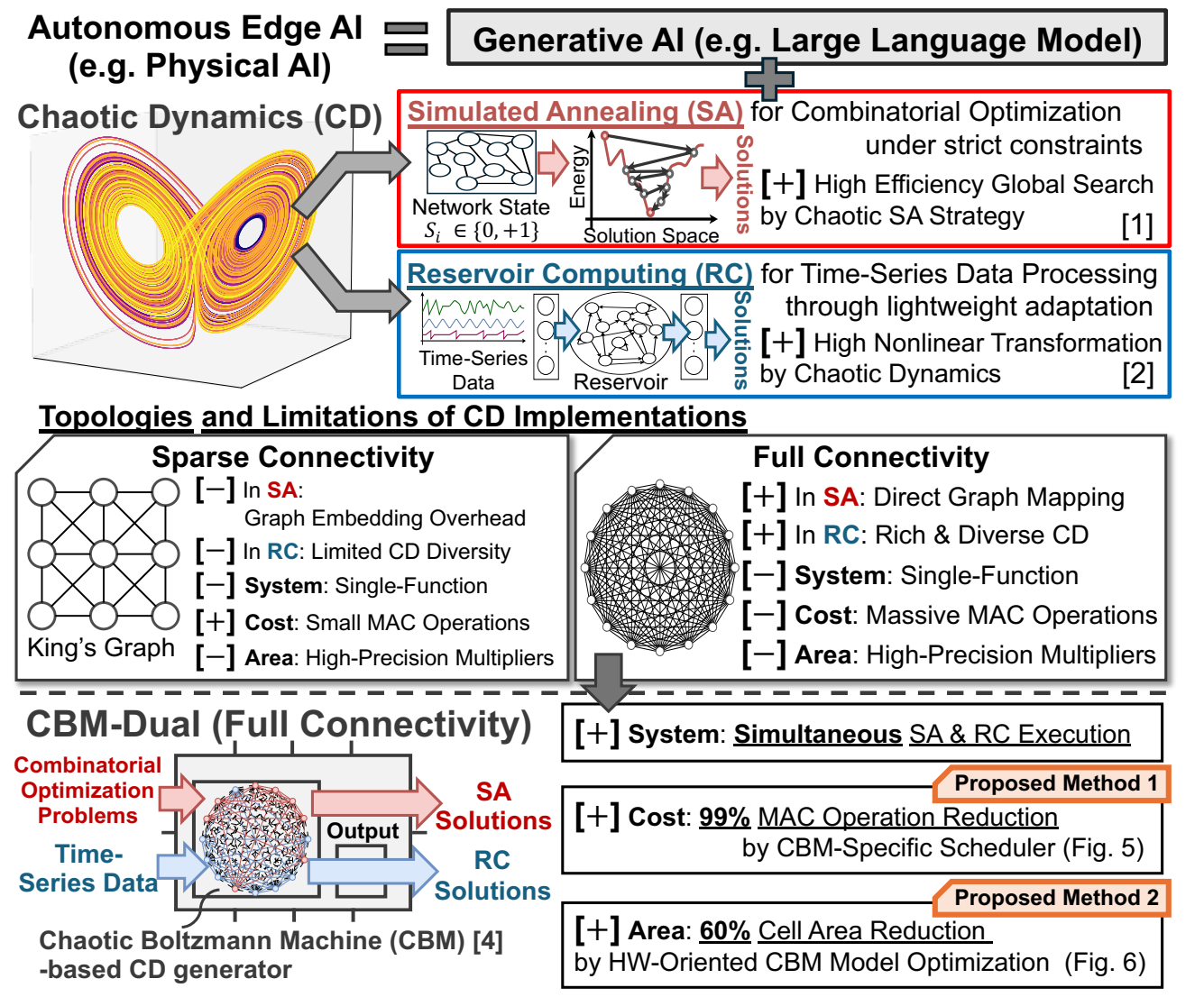

Fig.1. Concept of CBM-Dual complementing Generative AI for autonomous Physical AI and limitations of prior CD implementations.

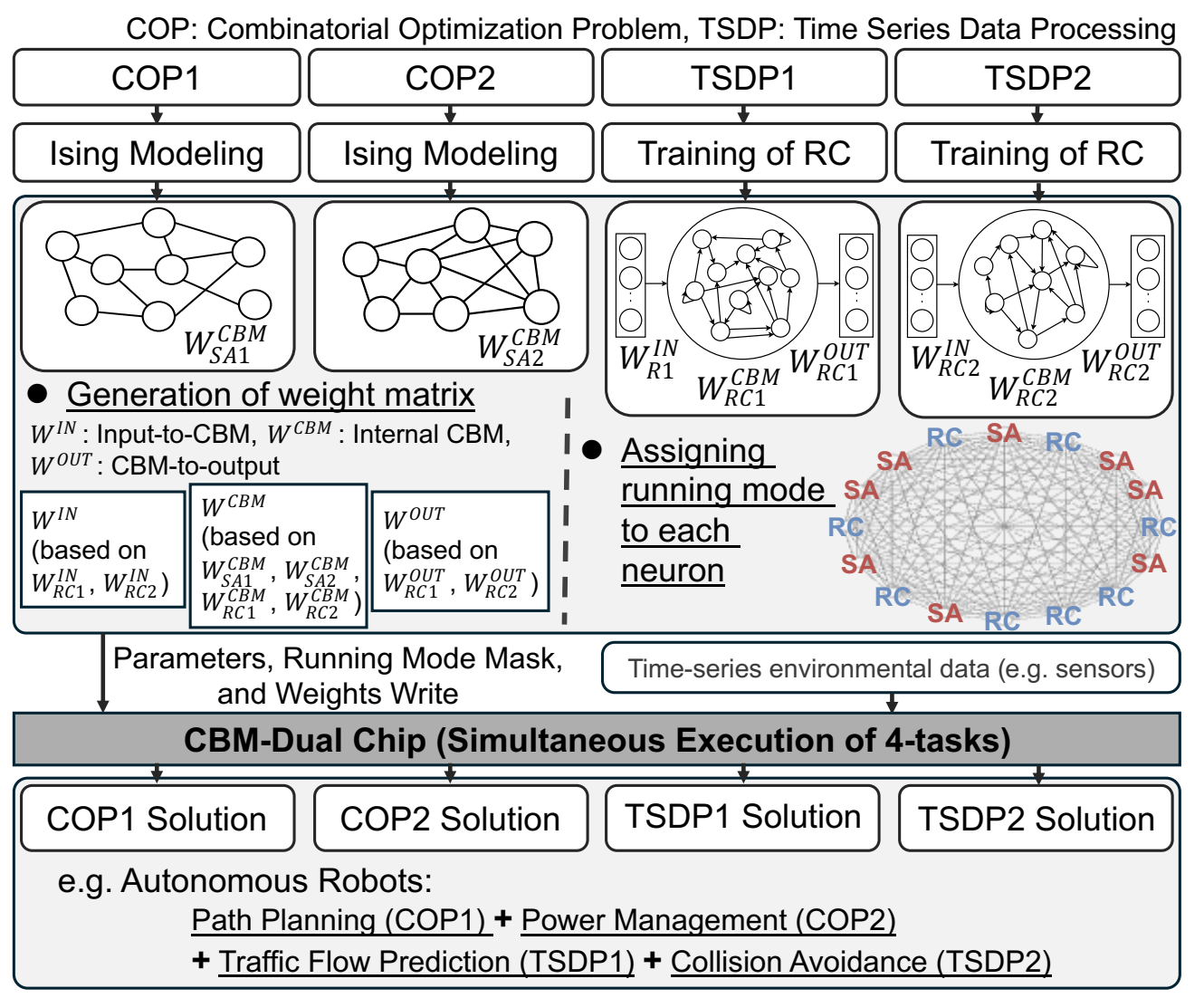

Fig. 2. Overall workflow of CBM-Dual: simultaneous execution of two combinatorial optimization and two time series data processing tasks.

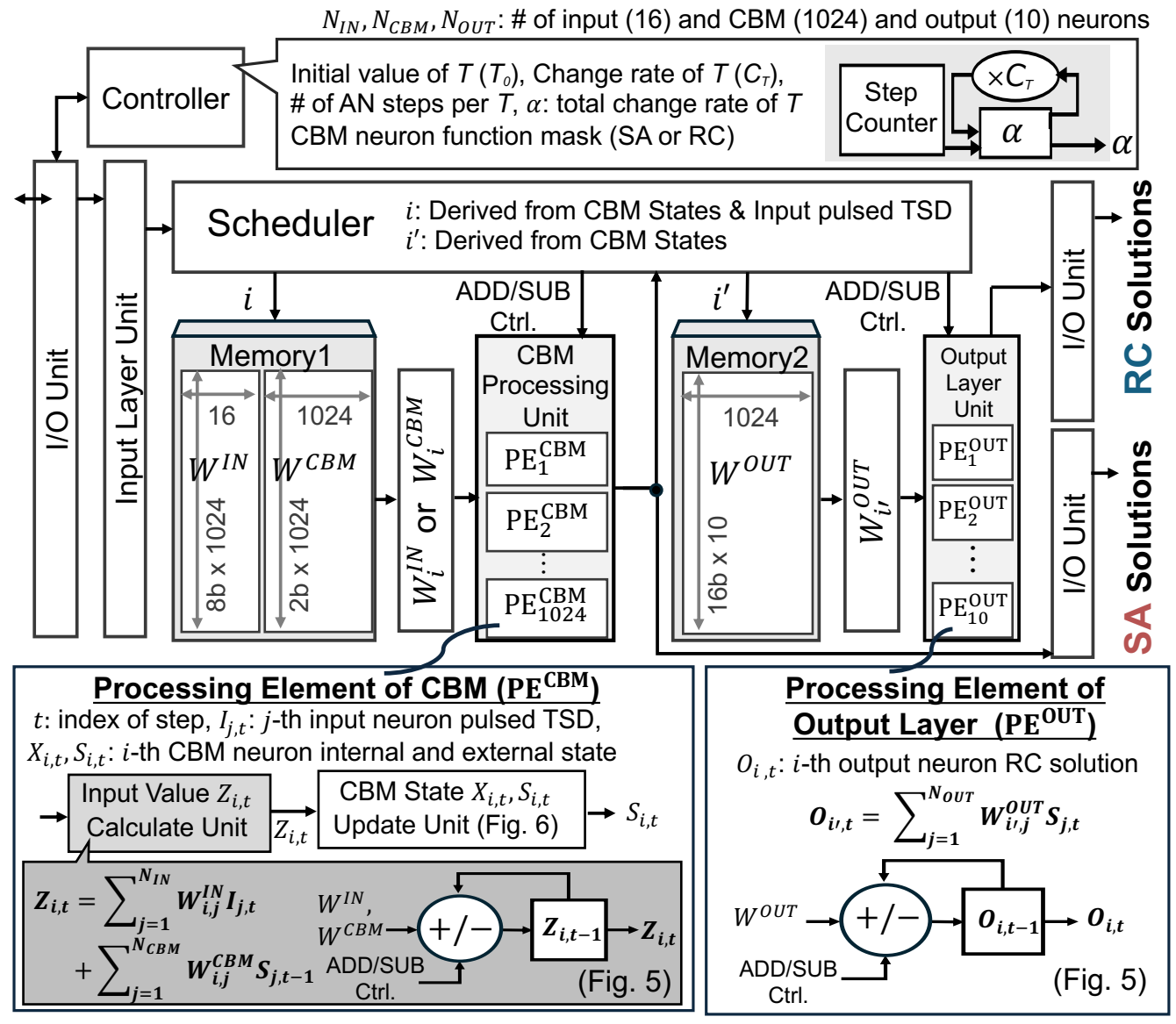

Fig. 3. Top-level CBM-Dual chip architecture and processing element design of CBM and output layer units.

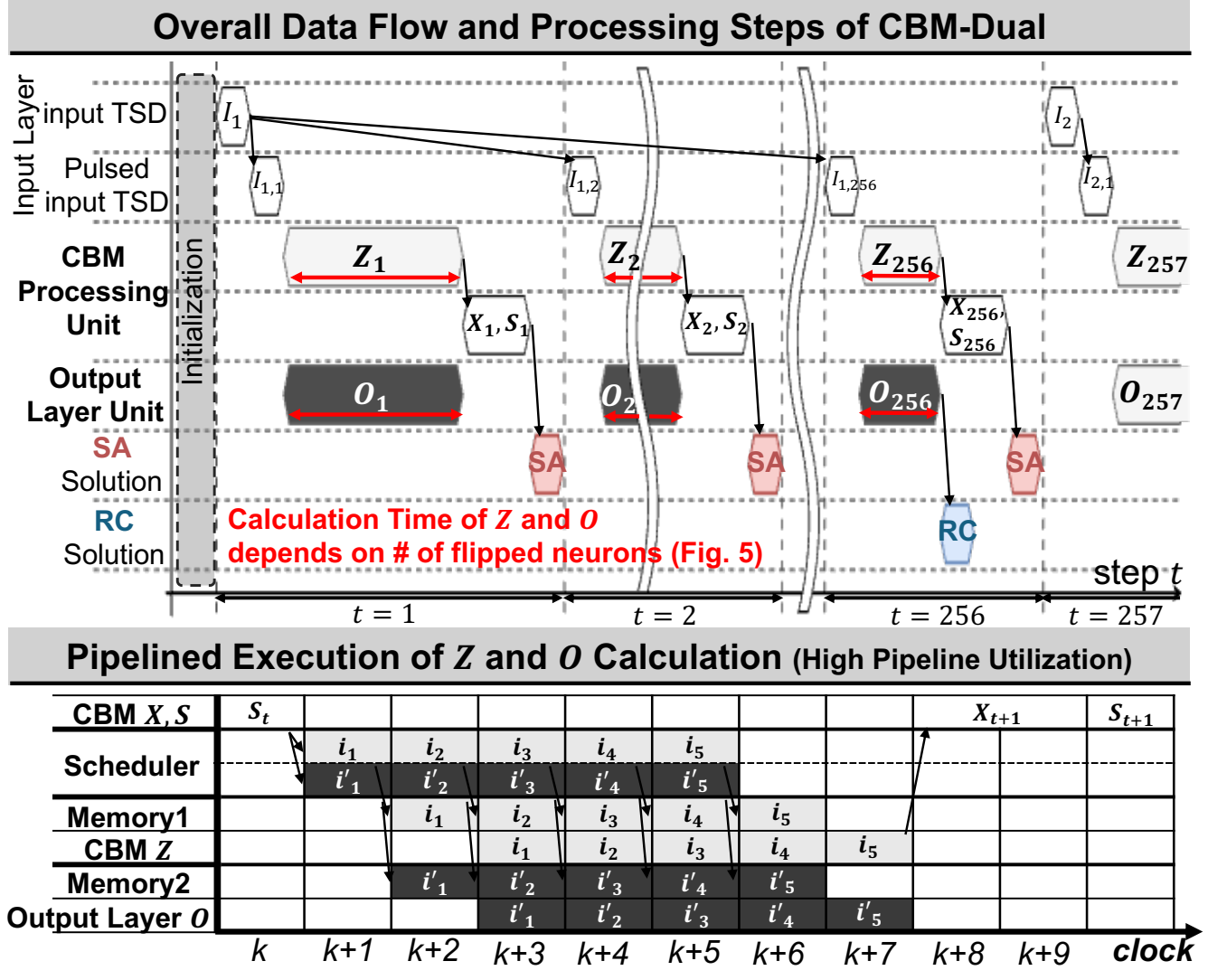

Fig. 4. Operation sequence of CBM-Dual: (top) macro-level data flow for dual-task processing, and (bottom) clock cycle-level pipelined timing diagram during MAC operation.

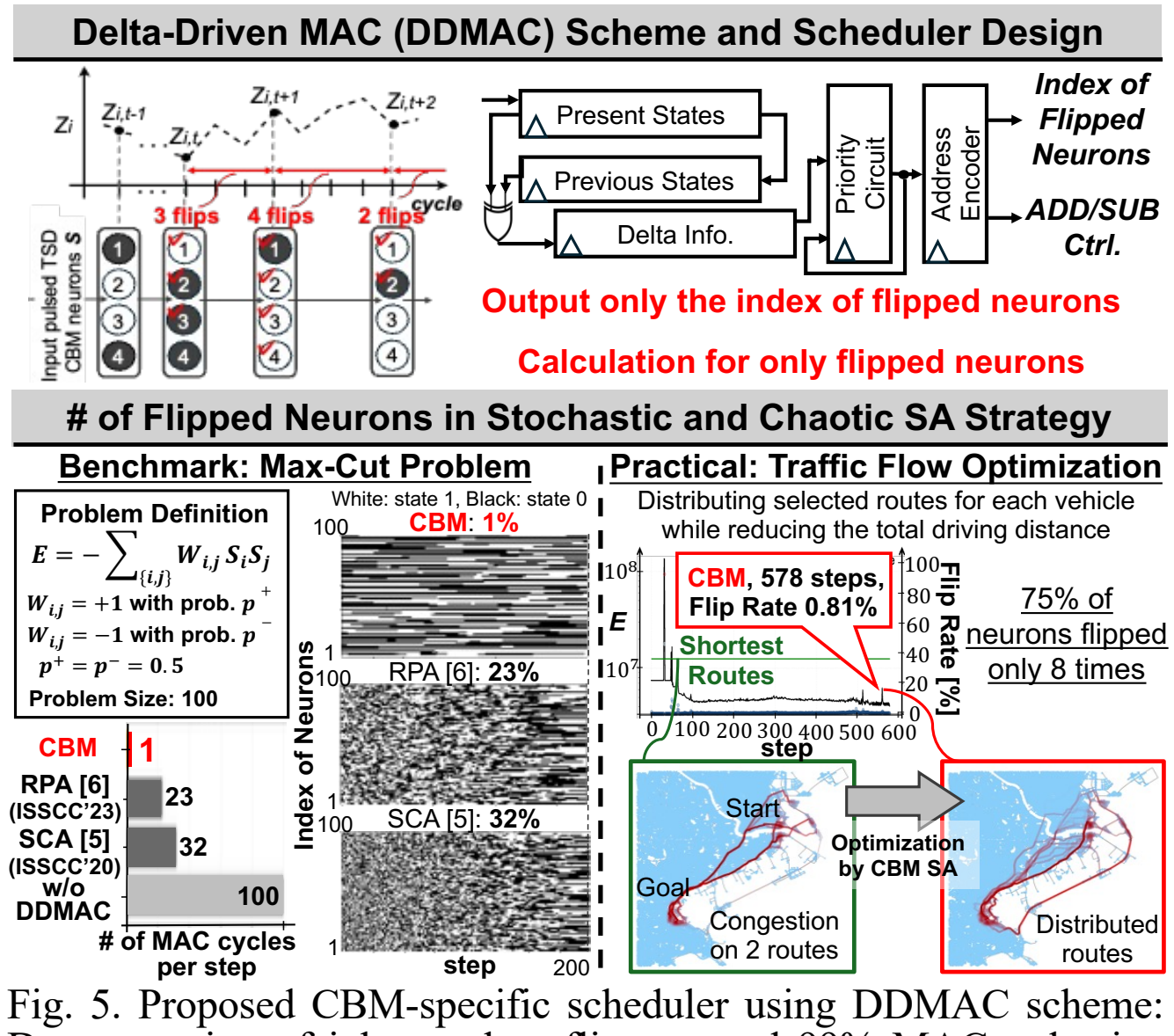

Fig. 5. Proposed CBM-specific scheduler using DDMAC scheme: Demonstration of inherent low flip rate and 99% MAC reduction with benchmark and practical problems.

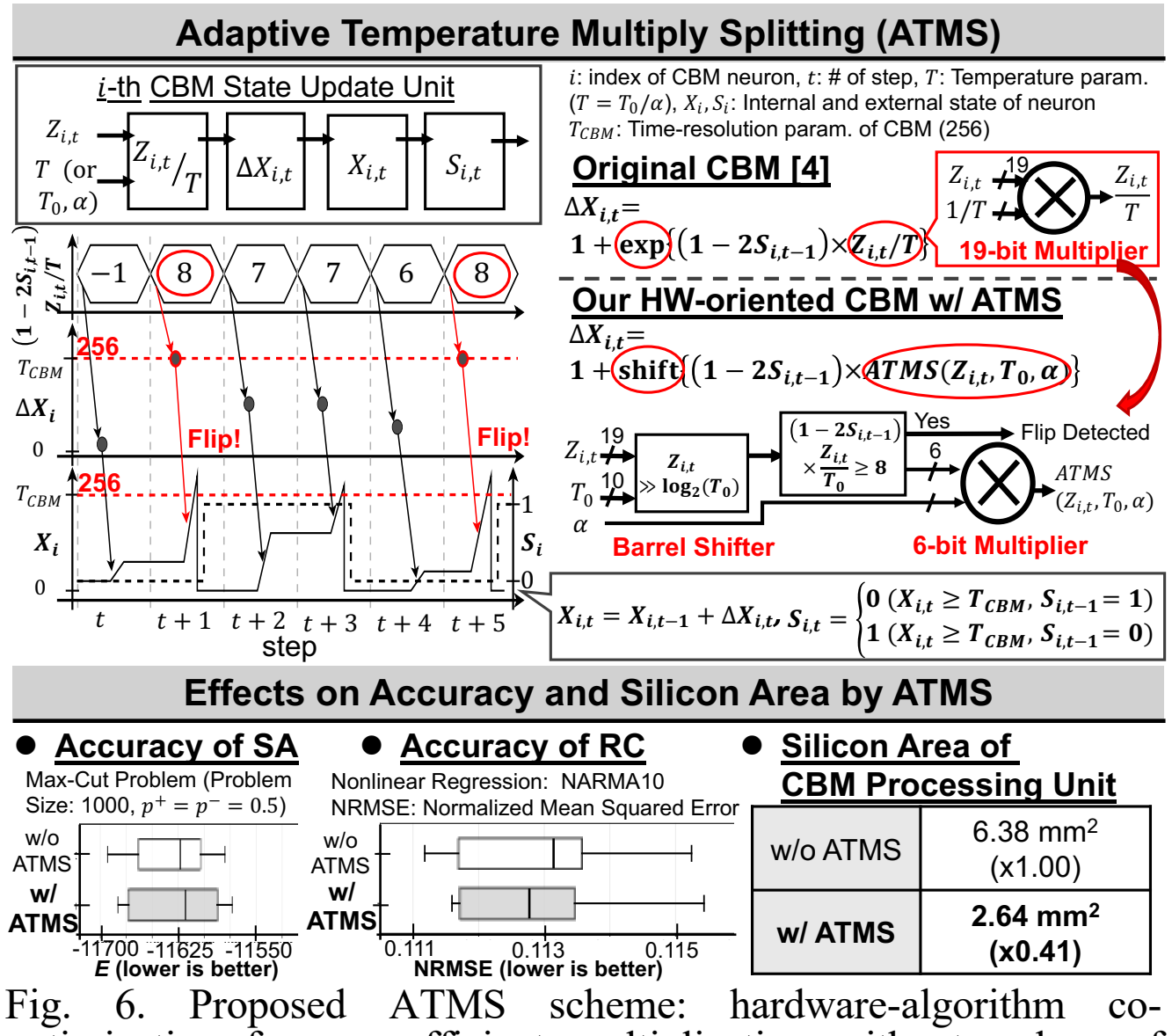

Fig. 6. Proposed ATMS scheme: hardware-algorithm co-optimization for area-efficient multiplication without a loss of accuracy in SA and RC tasks.

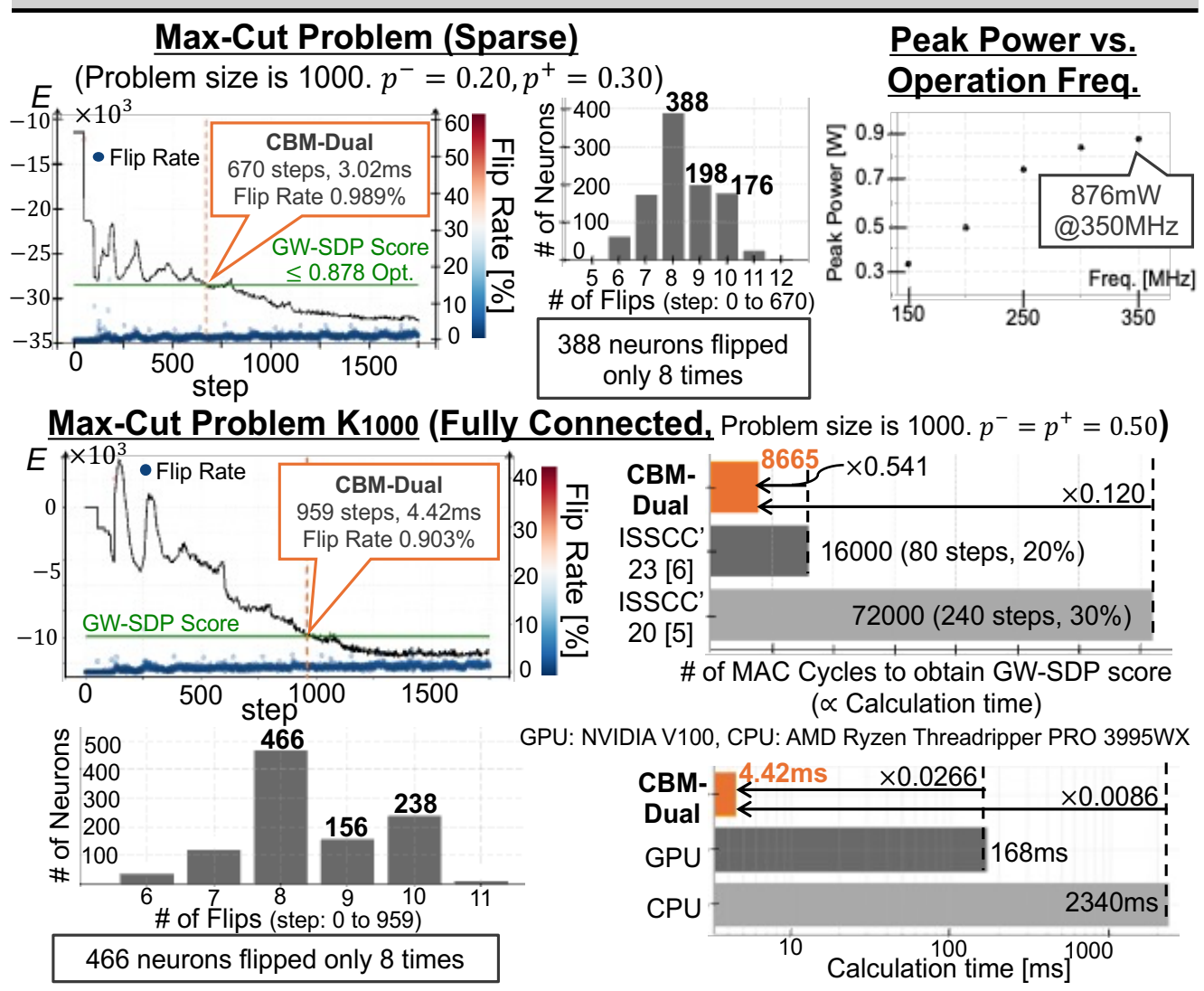

Fig. 7. Measurement results in SA mode: solving 1000-neuron max-cut problems and comparison with state-of-the-art SA processors.

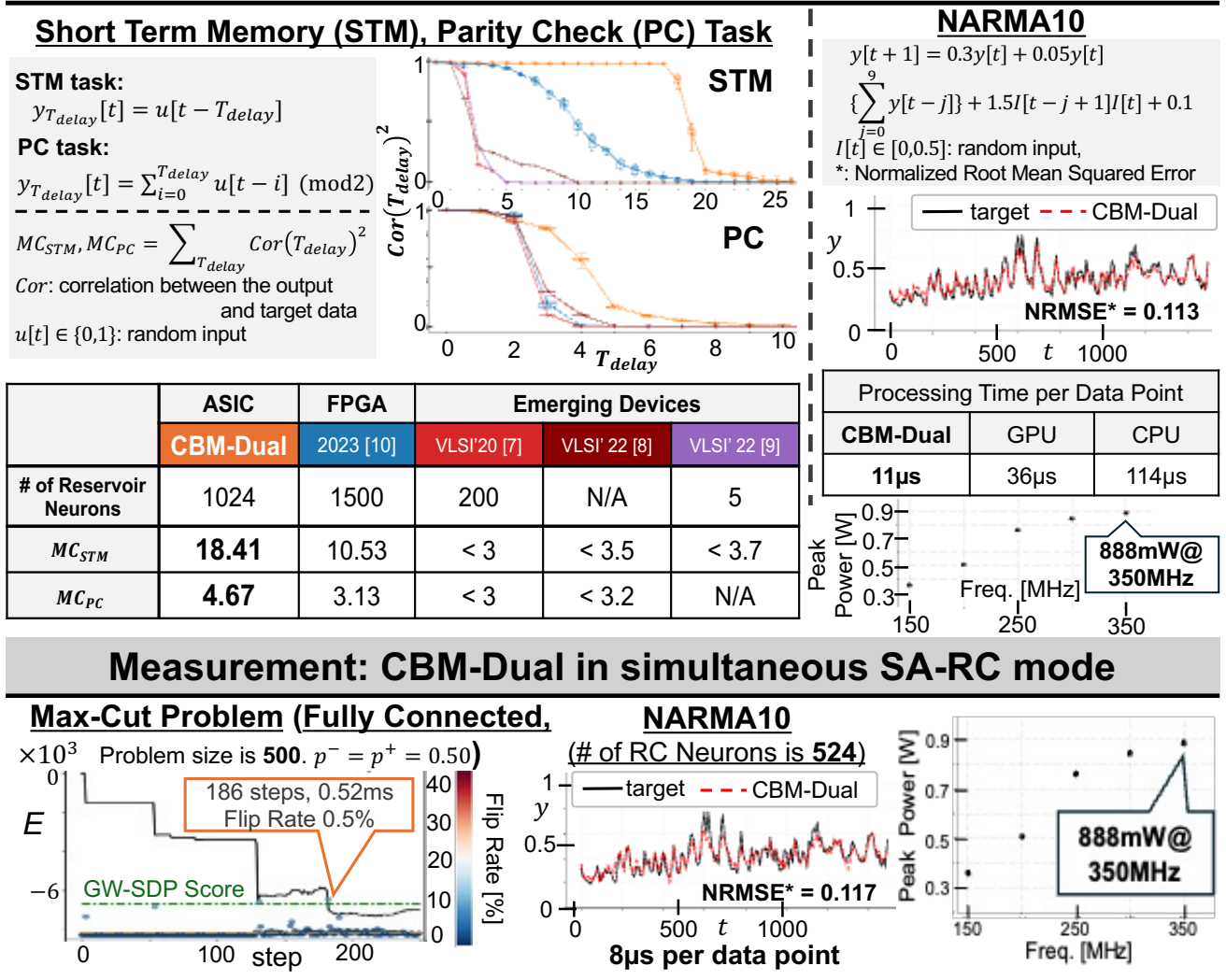

Fig. 8. Measurement results in (top) RC mode and (bottom) simultaneous SA-RC mode and comparison with state-of-the-art RC processors.

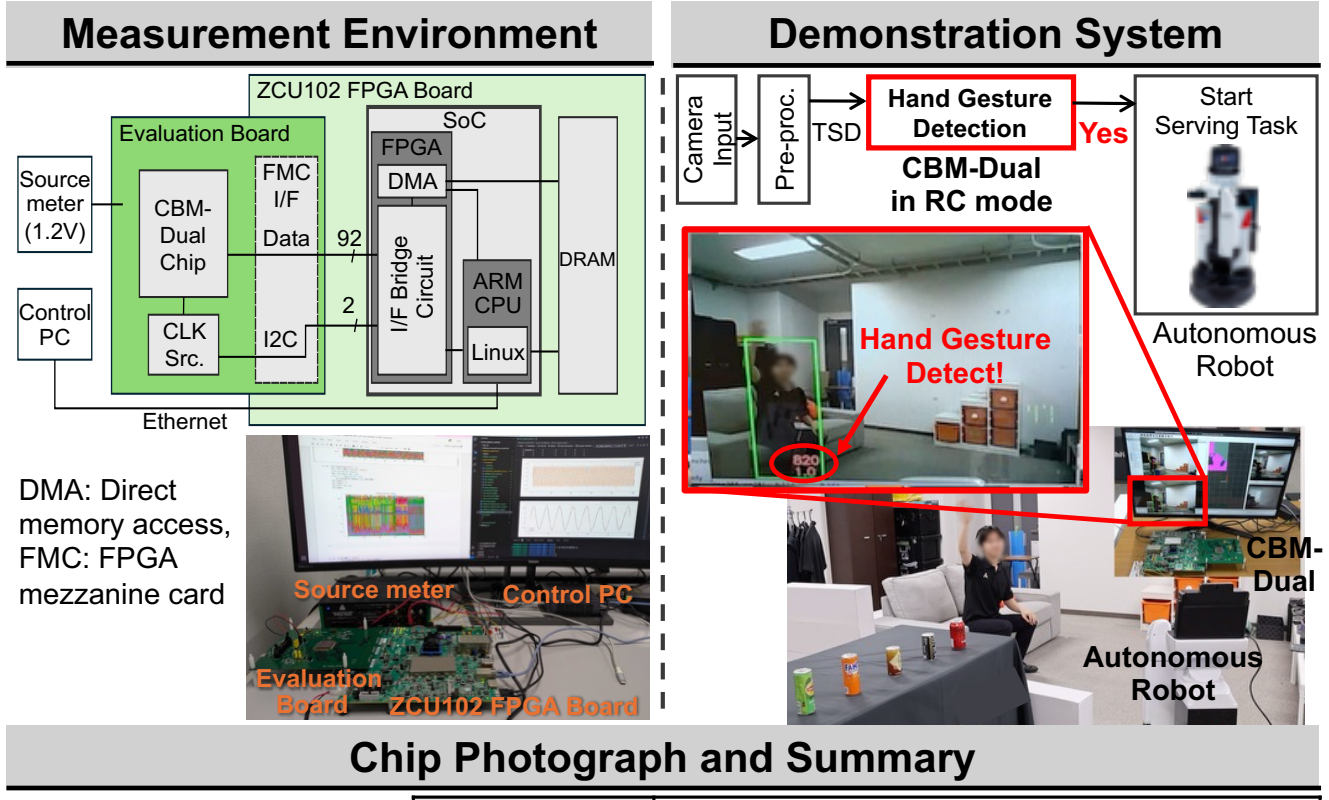

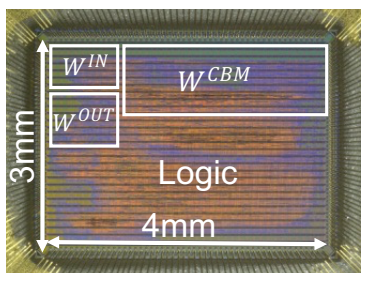

| | |
|---|---|
| Technology | 65nm CMOS (LP) |
| Chip Size | 3mm×4mm |
| Supply | Core: 1.2V, I/O: 1.8V |
| Max Frequency | 350MHz |
| Peak Power | 888mW |
| SRAM | 2.4Mb ($W^{IN}$: 0.13Mb, $W^{CBM}$: 2.1Mb, $W^{OUT}$: 0.16Mb) |

Fig. 9. (Top) measurement environment and application demonstration of hand gesture detection for autonomous robots using the RC function. (Bottom) chip photograph and summary.

TABLE I. Comparison with state-of-the-art ASIC SA processors.

| | ISSCC '20 [5] | ISSCC '23 [6] | ISSCC '24 [11] | This work |
|---|---|---|---|---|
| Technology | 65nm GP | 40nm LP | 40nm | 65nm LP |
| Topology | Fully Connected | | | |
| SA Algorithm | Stochastic (SCA, RPA, SA) | | | Chaotic |
| # of Neurons | 512 | 512 | 1024 | 1024 |
| Weight Precision | 5 | 8 | 6 | 2 |
| Op. Freq. [MHz] | 320 | 336 | 200 | 350 |
| Power [mW] | 649 | 475 | 110 | 876 |
| Solvable K1000 * | No | No | No | Yes |
| MACs / Step * ↓ | 300 | 200 | 1000 | 10 |
| Energy / Step / Neuron * [nJ] ↓ | 1.16 (×54) | 0.54 (×25) | 0.53 (×25) | 0.021 (×1) |
| Energy to GW-SDP Score * ↓ | 286 (×14) | 44 (×2) | N/A | 21 (×1) |

*: When Solving Max-Cut Problem K1000

TABLE II. Comparison with state-of-the-art RC processors.

| | VLSI '20 [7] | VLSI '22 [8] | VLSI '22 [9] | FPT '23 [10] | This work |
|---|---|---|---|---|---|
| Platform | Emerging Device | | | FPGA | CMOS 65nm LP |
| Topology | Sparse | | | | Fully Connected |
| Reservoir Size | 200 | N/A | 5 | 1500 | 1024 |
| Op. Freq. [MHz] | N/A | N/A | N/A | 100 | 350 |
| $MC_{STM}$ ↑ | < 3 | < 3.5 | < 3.7 | 10.5 | 18.4 |
| $MC_{PC}$ ↑ | < 3 | < 3.2 | N/A | 3.1 | 4.7 |
| NARMA10 *↓ | N/A | N/A | N/A | 0.133 | 0.113 |
| Latency / Point [µs]**↓ | N/A | N/A | N/A | 15 | 11 |
| Energy / Point [µJ]**↓ | N/A | N/A | N/A | 45 (×4.5) | 10 (×1.0) |

*: NRMSE, **: When Solving NARMA10